\documentclass[prb,preprint,showpacs]{revtex4}
\usepackage{graphicx}
\begin{document}

\title{Random lasing in weakly scattering systems}
\author{X. Wu$^{1}$, W. Fang$^{1}$, A. Yamilov$^{1,2}$, A. A. Chabanov$^{1,3}$, A. A. Asatryan$^{4}$, L. C. Botten$^{4}$, and H. Cao$^{1}$}
\affiliation{
$^{1}$ Department of Physics and Astronomy, Northwestern University, Evanston, Illinois, 60208, USA. \\
$^{2}$ Department of Physics, University of Missouri-Rolla, Rolla, Missouri, 65409, USA. \\
$^{3}$ Department of Physics and Astronomy, University of Texas-San Antonio, San Antonio, Texas, 78249, USA. \\
$^{4}$ Centre for Ultrahigh-bandwidth Devices for Optical Systems and Department of Mathematical Sciences, University of Technology, Sydney, New South Wales 2007, Australia.}

\begin{abstract}
We present detailed experimental and numerical studies of random lasing in weakly scattering systems. The interference of scattered light, which is weak in the passive systems, is greatly enhanced in the presence of high gain, providing coherent and resonant feedback for lasing. The lasing modes are confined in the vicinity of the pumped volume due to absorption of emitted light outside it. In the ballistic regime where the size of gain volume is less than the scattering mean free path, lasing oscillation occurs along the direction in which the gain volume is most extended, producing directional laser output. The feedback for lasing originates mainly from backscattering of particles near the boundaries of pumped region. It results in nearly constant frequency spacing of lasing modes, which scales inversely with the maximum dimension of the gain volume.   
\end{abstract}
\pacs{42.55.Zz,42.25.Dd}
\maketitle

\section{Introduction}

Random laser represents a non-conventional laser whose feedback is mediated by random fluctuation of dielectric constant in space. Since the pioneering work of Letokhov and coworkers \cite{letokhov}, lasing in disordered media has been a subject of intense theoretical and experimental studies \cite{cao_WRM}. Random lasers have been realized in various material systems, from semiconductor nanoparticles, ceramic powder to polymers, organic materials and biological tissues. Their low fabrication cost, sample specific lasing frequency, small size, flexible shape, and substrate compatibility lead to many potential applications \cite{cao_WRM}.

There are two kinds of feedback for random lasing: one is {\it intensity} or {\it energy} feedback, the other is {\it field} or {\it amplitude} feedback \cite{cao_WRM}. The field feedback is phase sensitive (i.e. coherent), and therefore frequency dependent (i.e. resonant). The intensity feedback is phase insensitive (i.e. incoherent) and frequency independent (i.e. non-resonant).  Based on the different feedback mechanisms, random lasers are classified into two categories: (i) random laser with incoherent and non-resonant feedback, (ii) random laser with coherent and resonant feedback. 

In a strongly scattering system, multiple scattering facilitates light (of wavelength $\lambda$) to return to the same coherence volume ($\sim \lambda^3$) it has visited before, providing field feedback for lasing. The lasing frequencies are determined by the interference of scattered light returning via different paths. However, lasing with field feedback was realized also in the weak scattering regime \cite{frolov,frolov1,frolov2,ling,polson,anni,mujumdar,polson1}. In 1999, Frolov and coworkers reported laser-like emission from several weakly scattering samples including the $\pi$-conjugated polymer films,  organic dyes-doped gel films, opal crystals saturated with polymer and laser dye solutions under strip pumping \cite{frolov,frolov1,frolov2}. They observed two pump thresholds, the first corresponding to spectral narrowing of amplified  spontaneous emission (ASE), and the second to the transformation of featureless ASE spectrum into a finely structured spectrum having numerous randomly positioned narrow lines. The fine spectral structures, which were reproducible under constant pumping condition and sample position, resulted from lasing with resonant feedback. The weak scattering, occurring repeatedly along the entire length of the excitation strip, supplied the distributed feedback for lasing. In 2004 Mujumdar and coworkers reported narrow emission spikes from suspensions of ZnO particles in  Rhodmaine 6G-methanol solutions over a broad range of scattering strengths\cite{mujumdar}. The spikes were distinct from shot to shot  and thus intrinsically stochastic. They were attributed to amplification of spontaneous emission along very long trajectories instead of optical cavities. Recently Polson and Vardeny used the power Fourier transform technique \cite{polson} to reveal the underlying periodicity of the emission peaks in several disordered organic gain media in the  weak scattering regime \cite{polson1}. The ensemble-averaged power Fourier transform of random laser emission spectra contained a sharp, well-resolved Fourier component and its harmonics, which was characteristic of a well-defined laser resonator. However, it was not clear how such resonator was formed. 

The above puzzles motivated us to study random lasing in the weak scattering regime. The main questions we intend to address in this papers are: (i) is the interference effect still important to random lasing in the weak scattering regime? (ii) what are the lasing modes in the  weakly scattering systems and how are they formed? To answer these questions, we performed  both  experimental studies and numerical simulations. 

The paper is organized as follows. A detailed experimental study of lasing in weakly scattering systems is presented in section II, followed by the numerical simulation in Section III. Section IV has a brief discussion and conclusion. 

\section{Experiments}

We performed experiments on several weakly-scattering systems which consisted of passive scatterers embedded in active homogeneous media. The scatterers included TiO$_{2}$ particles of radius 200 nm, ZnO particles of radius 38 nm, and SiO$_2$ particles of radius 220 nm. The nanoparticles were suspended in a laser dye solution, e.g., Rhodamine 640 perchlorate, stilbene 420 or LDS 722 in diethylene glycol (DEG) or methanol. The experimental results obtained with different particles, dyes and solvents were qualitatively similar. As an example, we will demonstrate the lasing phenomena with  colloidal suspensions of TiO$_{2}$ particles in DEG with Rhodamine 640. 

A small amount of TiO$_{2}$ (rutile) particles, with an average radius of 200 nm, were dissolved in the DEG solution of rhodamine 640 perchlorate dye. To prevent flocculation, the TiO$_{2}$ particles were coated with a thin layer of Al$_{2}$O$_{3}$. DEG was chosen as the solvent instead of the widely-used methanol because of the facts that (i) the windows of the quartz cuvette that contained the methanol solution were coated with a layer of TiO$_{2}$ particles, whereas such coating was not observed for the DEG solution;  (ii) the viscosity of DEG was about 30 times larger than methanol, thus the sedimentation of TiO$_{2}$ particles in DEG was much slower. In our experiment, the particle density $\rho$ ranged from $1.87 \times 10^{8}$ cm$^{-3}$ to $5.6 \times 10^{10}$ cm$^{-3}$. The scattering mean free path was estimated by $ l_s = 1/ \rho \sigma_s$, where $\sigma_s$ is the scattering cross section of a TiO$_{2}$ spherical particle with radius 200 nm. The value of $l_s$ varied from 1.07 cm to 35 $\mu$m. The dye molarity $M$ also changed from 3 to 10 mM. Right before the lasing experiment, the suspension was placed in an ultrasonic bath for 30 minutes to prevent sedimentation of the particles. During the experiment, the solution was contained in a quartz cuvette that was 1.0 cm long, 1.0 cm wide and 4.5 cm high. The dye molecules in the solution  were optically pumped by the frequency-doubled output ($\lambda_p$ = 532 nm) of a mode-locked Nd:YAG laser (25 ps pulse width, 10 Hz repetition rate). The pump beam was focused by a lens into the solution through the front window of the cuvette. The radius of the pump spot at the entrance to the solution was about 20 $\mu$m. The experimental setup is shown schematically in the inset of Fig. 1. The emission from the solution was collected in the backward direction of the incident pump beam. A second lens focused the emission into a fiber bundle (FB) which was connected to the entrance slit of a spectrometer with cooled CCD array detector. The spectral resolution was 0.6 $\AA$. 

We started the experiment with a sample of $M$ = 5 mM and $\rho = 3.0\times10^{9}$ cm$^{-3}$. At low pumping level, the emission spectrum featured the broad spontaneous emission band of rhodamine 640 molecules. Above a threshold pump intensity, discrete narrow peaks emerged in the emission spectrum, and their intensities grew rapidly with increasing pumping. This behavior corresponded to the onset of lasing. The lasing peaks could be as narrow as 0.12 nm. Their frequencies changed from pulse to pulse (shot). We repeated the  experiment with samples of different particle density but the same dye concentration. Lasing was observed only within certain range of particle density. Figure 1 show the spectra of emission from five samples taken at the same incident pump pulse energy 0.4 $\mu$J. Each spectrum was integrated over 25 shots. Only a relatively broad amplified spontaneous emission (ASE) peak was observed for the neat dye solution, whereas a few discrete lasing peaks emerged on top of the ASE spectrum at small particle concentration $\rho=1.87\times10^{8}$ cm$^{-3}$. Increasing particle density to $1.87\times10^{9}$ cm$^{-3}$ led to an increase in the number of lasing peaks and the peak intensity.  However, when $\rho$ increased further to $1.3\times10^{10}$ cm$^{-3}$,  the lasing emission started to decrease. Eventually at $\rho=5.0\times10^{10}$ cm$^{-3}$ lasing peaks disappeared. A further increase of the incident pump pulse energy to 1.2 $\mu$J resulted in an ASE peak at longer wavelength, shown in the right inset of Fig. 1. The red shift of the ASE peak might be caused by the surface effect on emission frequency of Rhodamine 640 molecules adsorbed on the TiO$_2$ particles. One support for this explanation was that the emission frequency was blue shifted when we replaced the TiO$_2$ particles by SiO$_2$ particles. 

In Fig. 2, the incident pump pulse energy at the lasing threshold $P_t$ is plotted against the particle density $\rho$.  At the lasing threshold, the slope of emission intensity versus pump pulse energy exhibited a sudden increase (inset of Fig. 2). 
At $\rho = 3.8 \times 10^{8}$ cm$^{-3}$, lasing started at 0.21 $\mu$J. 
As $\rho$ increased to $1.5 \times 10^{9}$ cm$^{-3}$, $P_t$ decreased gradually to 0.12 $\mu$J. Then it remained nearly constant with a further increase of $\rho$.  
The threshold started to rise at $\rho=1.9 \times 10^{10}$ cm$^{-3}$, then went up quickly with $\rho$. At $\rho=5.6 \times 10^{10}$ cm$^{-3}$, no lasing peaks were observed up to the maximum pump pulse energy 2.0 $\mu$J we used, although at 1.0 $\mu$J an ASE peak appeared at longer wavelength.

The particles played an essential role in the lasing process in our suspensions because lasing did not happen in the neat dye solution. One possibility was lasing within individual particles that served as laser resonators. It contradicted two experimental observations: (i) the lasing threshold depended on the particle density (Fig. 2); (ii) the laser output was highly directional (shown next).  The left inset of Fig. 3 is a sketch of our directionality measurement setup. A fiber bundle was placed at the focal plane of the lens. It was scanned with fine steps parallel to the focal plane. At each step, the spectrum of emission into a particular direction was recorded.  The output angle $\theta$ was computed from the fiber bundle position. Its range was limited by the diameter of the lens to about 14 degree. $\theta=0$ corresponded to the backward direction of the incident pump beam.  In each spectrum, the emission intensity was integrated over the wavelength range of 604 $-$ 612 nm in which the lasing peaks were located. Figure 3 is a plot of the integrated emission intensity versus the output angle $\theta$ at $\rho = 3.0 \times 10^9$ cm$^{-3}$. Although the spontaneous emission at low pumping was isotropic, the lasing emission was strongly confined to the backward direction of the incident pump beam. The divergence angle $\Delta \theta$ of the output laser beam was merely 4 degree. To check the effect of reflection by the front window of the cuvette on lasing, we rotated the cuvette around the vertical axis and repeated the measurement. As shown in the right inset of Fig. 3, $\phi$ represented the angle between the incident pump beam and the normal of the front window.  Similar lasing phenomena were observed except a small increase of the lasing threshold. The lasing emission was always confined to the backward direction of pump beam even when $\phi$ was much larger than the divergence angle of the focused pump beam, which was about 4 degrees. This result demonstrated that the front window of the cuvette was not indispensable to the lasing process. Figure 3 also shows the angular distribution of ASE from a sample of higher particle density ($\rho = 5 \times 10^{10}$ cm$^{-3}$). The integrated intensity of ASE (at longer wavelength) was nearly constant over the angular range of detection.   

To understand the directionality of lasing emission from the dilute suspension, the pumped region was imaged through a side window of the cuvette. The measurement setup is sketched in the inset of Fig. 4(a). Emission from the excited region was collected through the side window by a 5$\times$ objective lens and imaged onto a CCD camera by integrating multiple pulses. Spectrum was taken simultaneously by partitioning the signal with a beam splitter (BS). Figure 4(a) compares the spectrum of emission through the side window to that through the front window of the cuvette from the same sample ($\rho = 3 \times 10^9$ cm$^{-3}$, $M$ = 5 mM) under identical pumping condition. The spectrum of emission from the front window exhibited large lasing peaks. However, only spontaneous emission was observed through the side window, and it shifted to longer wavelength as a result of reabsorption in the unpumped solution between the excited region and side window. To calibrate the reabsorption, we measured the spontaneous emission spectra at low pump intensity from both front and side windows. The magnitude of reabsorption was estimated from the intensity ratio of emission through the side window to that through the front. Based on this estimation, 
we concluded that the reabsorption was not strong enough to make the lasing peaks, which emerged in the front emission spectrum at high pump intensity, disappear in the side emission spectrum. This conclusion confirmed the result of lasing directionality measurement in Fig. 3.     
More importantly, the image of spontaneous emission intensity distribution taken through the side window exhibited the shape of the excited region in the sample. As shown in Fig. 4(b), the excited volume at low particle density had a cone shape. The length of the cone was much larger than its base diameter. Unfortunately, we could not get the exact length of the cone from the image, because near its end the spontaneous emission was too weak to be recorded by the CCD camera. At high particle density, the shape of excited volume changed to hemisphere as shown in Fig. 4(c) at a higher pumping power. This change was caused by increased scattering of pump light. In Fig. 4(b) $\rho=3.0\times10^{9}$ cm$^{-3}$, the scattering mean free path $l_{s}$ at the pump wavelength $\lambda_p = 532$ nm was estimated to be 800 $\mu$m. The (linear) absorption length $l_a$, obtained from the transmission measurement of neat dye solution, was about 50 $\mu$m at $\lambda_p = 532$ nm. Strong pumping in the lasing experiment could saturate the absorption of dye molecules, leading to an increase of $l_a$. Since the shape of the excited volume shown in Fig. 4(b) was nearly identical to that in the neat dye solution, the scattering of pump light must be much weaker than absorption, i.e., $l_a$ was still shorter than $l_s$. In Fig. 4(c) $\rho=5.0\times10^{10}$ cm$^{-3}$, $l_{s}$ was shortened to 53 $\mu$m. Scattering of pump light became much stronger. As a result of multiple scattering, the cone was replaced by a hemisphere. The image of excited volume provided some clue to high directionality of lasing emission at small $\rho$ and non-directionality of ASE at large $\rho$ in Fig. 3. At low particle density, stimulated emission in the cone-shaped gain volume was the strongest along the cone due to the longest path length. Since the cone was parallel to the incident pump beam, lasing was confined to the direction parallel to the pump beam. The divergence angle $\Delta \theta$ of laser output was determined by the aspect ratio of the excited cone, namely, $\Delta \theta \sim 2 r_p/ L_p$, where $L_p$ is the cone length and $r_p$ is the base radius. At large $\rho$, emitted photons experienced multiple scattering while being amplified in the hemisphere-shaped gain volume. Hence, the ASE was nearly isotropic. 

Therefore, the shape of gain volume determined the lasing directionality, i.e., lasing occurred along the direction in which the gain volume was most extended. However, it was still not clear how the laser cavities were formed in the dilute suspension. We examined the lasing spectra more carefully by taking single-shot emission spectra with the setup shown in the inset of Fig. 1. Surprisingly, in most single-shot spectra the spectral spacing of the lasing peaks was close to a constant.  Figure 5(a) is an example of single-shot emission spectrum taken from the sample of $\rho = 1.87\times10^{9}$ cm$^{-3}$ and $M$ = 5 mM. The spectral correlation function $C(d \lambda) \equiv \langle I(\lambda) I (\lambda + d \lambda) \rangle / \langle I(\lambda)^2 \rangle$ was computed for the spectrum in Fig. 5(a) and plotted in Fig. 5(b). The almost regularly spaced correlation peaks revealed the periodicity of lasing peaks. Despite that the lasing peaks completely changed from shot to shot, the peak spacing was nearly the same. In the spectrum taken over many shots, the periodicity was smeared out due to random (uncorrelated) peak positions in different shots. As an example, in Fig. 1 the solid thick curve represents the lasing spectrum integrated over 25 pulses, and the lasing peaks did not exhibit clear periodicity. We also noticed that the periodicity was less obvious at higher particle density.

We would like to point out that the lasing peaks in the dilute suspension of particles are fundamentally different from the stochastic ASE spikes that could be observed also in the neat dye solution. Figure 6 shows a single-shot spectrum of emission from the DEG solution of 5 mM Rhodamine 640 without any particles. The spectrum was taken under the same condition as that in Fig. 5. At high pumping level, stochastic spikes appeared on top of the ASE peak. The spikes in Fig. 6 were denser and narrower than the lasing peaks in Fig. 5. They changed constantly from shot to shot. When integrating the spectrum over subsequent shots, the spikes were quickly averaged out, leaving a smooth ASE spectrum shown as the dotted curve in Fig. 1. Note that the stochastic spikes also appeared in the spectrum of emission from the dye solutions with particles. However, they were taken over by the huge lasing peaks at high pumping level. The stochastic structure of the pulsed ASE spectrum was first reported thirty years ago \cite{korolev,ishchenko74}. Since then, there have been detailed experimental and theoretical studies on this phenomenon \cite{korolev76,brazovsky79,bor,sperber}. The spectral fluctuation originated from random spontaneous emission, which was strongly amplified as it propagated through the cone-shaped pump volume.  Non-stationary interference of the partially coherent ASE not only presented a grainy spatial pattern, but also caused drastic temporal fluctuation of intensity. The random intensity fluctuation within an ASE pulse in the time domain generated stochastic spikes in the spectral domain. According to the Fourier transformation, the width of spectral spikes 
was inversely proportional to the ASE pulse duration. In our case of picosecond pumping, the ASE pulse duration was of the order 25 ps. Thus the average spike width should be $\sim$ 0.05 nm, which was close to the measured value of 0.07 nm. 

To find out the location and size of the laser cavities in the dilute suspension, we placed a metallic rod in between the excited volume and the back window (inset of Fig. 7). On one hand, the rod prevented the emission from being reflected by the back window into the gain volume. On the other hand, a Fabry-Perot cavity was formed. Lasing in this cavity produced equally spaced peaks in the emission spectrum. The peak spacing $\Delta \lambda$ was determined by the cavity length $d$ (the distance between the rod and the front window), $\Delta \lambda = \lambda^2/2 n_e d$, where $n_e$ is the effective index of suspension.  As plotted in Fig. 7, $\Delta \lambda$ decreased with increasing   $d$. However, when $d$ exceeded a critical value $d_0 \simeq$ 450 $\mu$m, $\Delta \lambda$ jumped to a constant value and did not change with $d$ any more (Fig. 7). The lasing spectrum and peak spacing at $d > d_0$ were identical to those from the same sample without the metallic rod. This result demonstrated that the laser cavity in the colloidal solution was located within 450 $\mu$m from the front window, i.e., in the vicinity of excited cone. Despite the scattering mean free path  $l_s$ ($\sim 800$ $\mu$m) was much longer than the wavelength, the laser cavity was not extended over the entire sample, but confined to a region of dimension less than $l_s$. This could be explained by the reabsorption of laser emission in the unpumped part of sample. Our white-light absorption measurement and photoluminescence measurement of neat dye solution showed that Rhodamine 640 molecules in DEG had significant overlap between the absorption band and emission band. At the dye concentration $M$ = 5 mM, the absorption length $l_e$ at the emission wavelength $\lambda_e$ $\sim$ 610 nm was about 300 $\mu$m. At low density of TiO$_2$ particles in the solution, the absorption length of emitted photons in the unpumped region was shorter than the scattering length. If the emitted light traveled beyond one absorption length from the pumped region, its chance of returning to the pumped region was extremely low. Therefore, the reabsorption of emission suppressed the feedback from the unpumped region of the system, and effectively reduced the system (or cavity) size \cite{yamilov}. 

The above experiment illustrated the laser cavity was confined in the vicinity of the excited region. The directionality of laser output suggested the laser cavity was oriented along the excited cone. Nearly constant spectral spacing of lasing modes resembled that of a Fabry-Perot cavity. Using the formula of a Fabry-Perot cavity, we derived the cavity length $L_c$ from the average wavelength spacing $\Delta \lambda$ of the lasing peaks, $L_c = \lambda^2/2 n_e \Delta \lambda \sim 200 - 300$ $\mu$m. The estimated cavity length was close to the length of excited cone observed from the side images. The cone length was determined by the penetration depth $L_p$ of pump light into the suspension. When the scattering mean free path $l_{s}$ was much larger than the absorption length at the pump wavelength, $L_p$ was determined solely by absorption of pump light. This could explain the experimental observation that the averaging spacing of lasing peaks barely changed when the scattering mean free path $l_{s}$ was varied by more than one order of magnitude.  The cavity length $L_c$ (or the penetration depth $L_p$) did not depend on $l_{s}$ as long as $l_s$ exceeded the absorption length. This result was confirmed by the side images of excited cones. 

In the case of linear absorption, $L_p$ should be on the order of the (linear) absorption length $l_{a}$. However, the pumping in the lasing experiment was so intense that it saturated the absorption of dye molecules.  The saturation photon flux density $I_{s}=1/\sigma_{f}\tau_{f}$, where $\sigma_{f}$ is the fluorescence cross section and $\tau_{f}$ is the lifetime of dye molecules in the excited state. For Rhodamine 640 molecules, $\sigma_f$ is of the order $10^{-16}$ cm$^{2}$, and $\tau_{f}$ $10^{-9}$ s. Thus, $I_{s}$ $\sim 10^{25}$ cm$^{-2}$s$^{-1}$. The typical pump pulse energy at the lasing threshold was $\sim 0.1$ $\mu$J. From the pump pulse duration and pump spot size, we estimated that the incident pump photon flux density $I_{p} \sim 10^{27}$ cm$^{-2}$s$^{-1}$, which was two orders of magnitude higher than $I_{s}$. Hence, at the lasing threshold the absorption of dye molecules near the front window of the cuvette was strongly saturated, and the penetration depth $L_{p}$ was much longer than the linear absorption length $l_{a}$. 

To confirm $L_c \sim L_p$, we changed $L_p$ by varying the dye concentration. The higher the dye concentration, the shorter the penetration depth. If $L_c \sim L_p$, the spacing of lasing peaks $\Delta \lambda$ should increase.  Figure 8(a) shows the single shot lasing spectra from three solutions of $M$ = 3, 5, 10 mM. The particle density $\rho$ was fixed at $3\times 10^9$ cm$^{-3}$. It was evident that the spacing of lasing peaks increased at higher dye concentration. Images of the excited cones in the inset of Fig. 8(b) directly show that the excited cone was longer in the solution of lower dye concentration. Figure 8(b) plots the spectral correlation functions for the three spectra in Fig. 8(a). From them we extracted the average  wavelength spacing $\Delta \lambda$ = 0.34, 0.48, 0.96 nm for $M$ =  3, 5, 10 mM. This result confirmed that the laser cavity length was determined by the pump penetration depth. Hence, the laser cavity was located in the excited cone, and the cavity length was approximately equal to the cone length.


Although the lasing phenomenon in the weakly-scattering dye solution seemed to resemble that in a Fabry-Perot cavity with the ``mirrors'' at the tip and base of the excited cone, the question remained what made of the mirrors in the dilute colloidal suspension. The experiments presented earlier in this section already ruled out the front and back windows of the cuvette as the mirrors. The mirrors could not be formed by the nonlinear change of refractive index (both its real and imaginary parts) of the dye solution under intense pumping, otherwise lasing would have occurred also in the neat dye solution without particles. Another possibility is that the particles aggregated in the solution to form large clusters that served as mirrors. We monitored the solution during the experiment by imaging it onto a CCD camera through a side window of the cuvette. No clusters of size larger than 1$\mu$m were observed in the pumped solution. We removed the clusters smaller than 1 $\mu$m by filtering the solution. After the filtering, lasing phenomena remained the same. We also tried other suspensions such as ZnO in DEG and SiO$_2$ in methanol. Our previous studies confirmed the absence of particle aggregation in these suspensions, but we still observed similar lasing phenomenon. Therefore, the ``mirrors'' were not clusters of particles. Another candidate was bubbles or shock waves that were generated by the pump pulse \cite{chastov,bezrodnyi,vogel}. We indeed observed bubbles in the solution when the pump beam was very strong and its focal spot was close to the front window of the cuvette. When the bubbles were big enough to be seen with our imaging apparatus, lasing peaks disappeared. Hence, big bubbles did not facilitate lasing. Small bubbles, which were invisible, might be generated when the pumping was not very high. Such bubbles were usually generated at the focal spot of the pump beam where the pump intensity was the highest. Thus the bubble formation should be sensitive to the distance between the focal spot and the front window, which affected the pump intensity at the focal spot due to absorption in the solution. When we shifted the focal spot by moving the lens, the lasing behavior remained the same. This result eliminated the possibility of small bubbles contributing to lasing. All the experimental results led us to the conclusion that the coherent feedback for lasing came from the particles in the solution, more specifically, from the particles located near the tip and base of the excited cone. However, there were many particles inside the pumped volume, e.g., at $\rho=3 \times 10^9 $ cm$^{-3}$ the number of particles inside the excited cone was about 400. Why the feedback from the particles near the two ends of the cone dominated over that from the particles inside the cone? To understand this phenomena, we performed numerical calculations to be presented in the next section.

\section{Numerical simulation}

Several models were set up in the theoretical studies of stimulated emission in active random media, e.g. the diffusion equation with gain \cite{wiersma,john} and the Monte Carlo simulation \cite{berger,mujumdar}. These models calculated light intensity instead of electromagnetic field, thus they ignored the interference effect. Although it is usually weak in the diffusive regime, the interference effect is not always negligible. One example is the coherent backscattering, namely, the interference between the counter-propagating light enhances the backscattered intensity by a factor of two. The experimental results in the last section illustrated that the interference effect was significant in weakly scattering samples with gain, leading to lasing with resonant feedback. In the dilute suspensions of particles in which lasing could be realized, the gain volume had a cone shape and the cone length was shorter than the scattering mean free path. If we considered only one photon, it most likely would not be scattered as it traveled from one end of the cone to the other. However, the intense pumping generated a huge number of emitted photons. Despite the low probability of {\it one} photon being scattered, a significant number of emitted photons  were scattered by the particles in the excited cone. Some of them were scattered backwards, providing feedback for lasing along the cone. Such weak feedback was greatly amplified as the backscattered light propagated along the cone. The interference of the backscattered light determined the lasing frequencies.

Therefore, in the presence of large gain, the interference of scattered light is not negligible even in weakly scattering samples. To include the interference effect, we directly calculated the electromagnetic field in a random medium by solving the Maxwell's equations using the finite-difference time-domain (FDTD) method \cite{taflove}.  The optical gain was modeled as negative conductance 
\begin{equation}
\sigma(\omega)=-\frac{\sigma_0/2}{1+i(\omega-\omega_0)T_2}-\frac{\sigma_0/2}{1+i(\omega+\omega_0)T_2}.
\label{sigma}
\end{equation}
$\sigma_0$ determined the gain magnitude, $\omega_0$ and $1/T_2$ represented the center frequency and width of the gain spectrum, respectively.  We neglected the gain saturation, and limited our calculations to the regime just above the lasing threshold. 
A seed pulse with broad spectrum was launched at $t=0$ to initiate the amplification process.  The lasing threshold was defined by the minimum gain coefficient ($\sigma_0$) at which the electromagnetic field oscillation built up in time.

Our numerical calculations aimed not at reproducing the experimental results, but at addressing the key issues and providing physical insight to the lasing mechanism. Hence, we simulated lasing in two-dimensional systems in order to shorten the computing time. To model the elongated gain volume in the experiment, optical gain was introduced to a strip of length $L_p$ and width $W_p$. The refractive index was set at 1.0 both inside and outside the strip. Dielectric cylinders of radius 100 nm and refractive index 2.0 were introduced as scattering centers.  The strip dimension was much smaller than the experimental value due to the limited computing power. Consequently, the number of cylinders inside the strip was reduced to keep the system in weak scattering regime. 

The experimental results in the last section suggested that the coherent feedback for lasing resulted mainly from the particles located near the tip and base of the excited cone. In the dilute suspension of particles, there was probably only one particle located at the tip of the cone. One question was whether the backscattering of a single particle could provide enough feedback for lasing. To answer this question, we started with only two particles in the gain strip, one at each end, in the numerical simulation. The total system size was 16 $\mu$m $\times$ 8 $\mu$m, and the gain strip 8 $\mu$m $\times$ 4 $\mu$m.  At the system boundary, there was a perfectly matched absorbing layer. Two cylinders of radius 100 nm, placed at the two ends of the gain strip, had a separation of 8 $\mu$m. When the gain coefficient $\sigma_0$ was above a threshold value, we observed lasing oscillation. The spatial distribution of lasing intensity revealed that lasing occurred along the strip with the feedback from the two particles. The emission spectrum, obtained by Fourier transform of the electric field, consisted of multiple lasing modes equally spaced in frequency. We repeated the calculation after reducing the separation between the two particles inside the gain strip while keeping the strip length constant. The frequency spacing of lasing modes scaled inversely with the particle distance. These results confirmed lasing in the resonator composed of only two scatterers. 


In 1998, Wilhelmi proposed a laser composed of two Rayleigh scatterers with gain medium in between \cite{wilhelmi}.  We generalized the Rayleigh scatterers to Mie scatterers and derived the lasing threshold condition:
\begin{equation}
{{\sigma_b}\over{L_c^{m-1}}} e^{g_e L_c} = 1.
\label{threshold}
\end{equation}
$\sigma_b$ is the backscattering cross section of one particle, which depends on the particle size, refractive index and  light wavelength. The cavity length $L_c$ is the separation between the two scatterers. $m$ is the dimensionality of the scattering system. $g_e$ is the threshold gain coefficient for lasing. $\sigma_b/L_c^{m-1}$ describes the probability of a photon being backscattered by one particle and propagating to the other particle. The larger the $L_c$, the less percentage of the backscattered photons can reach the other particle. It seems to suggest that the quality factor of the two-particle cavity decreases with increasing $L_c$. This perception is incorrect. At the lasing threshold, the cavity loss is equal to the gain, namely, the loss per unit length $\alpha=g_{e}$. In the absence of intracavity absorption, $\alpha$ is related to the cold-cavity $Q$ as $\alpha=1/Q\lambda$. From the threshold gain coefficient $g_e$ in Eq.(\ref{threshold}), we derive the cold-cavity quality factor $Q=L_c/(\lambda |(m-1) \ln L_c - \ln \sigma_b|)$. As $\L_{c} \rightarrow \infty$, the numerator in the expression of $Q$ diverges faster than the denominator, thus $Q \rightarrow \infty$. The rise of $Q$ with $L_c$ is attributed to the increase of one-path length of light inside the cavity. The reduction  in the probability of photons backscattered by one particle then reaching the other particle in a long cavity is offset by the increase of photon travel time from one particle to the other.  Hence, the lasing threshold decreases with increasing $L_c$. The resonator with the lowest lasing threshold is composed of two particles with the largest possible separation inside the gain volume. 

Our numerical simulations also demonstrated directional lasing output from the two-scatterer cavity. The near-field to far-field transformation of electric field gave the output laser intensity as a function of polar angle. Figure 9 shows the numerical data for three gain strips of length $L_p = 4, 8, 16$ $\mu$m. The strip width $W_p$ was fixed at 4 $\mu$m. The two scatterers were always placed at the ends of the strip. From the envelop of the far-field intensity distribution, we obtained the angular width of  the output laser beam. It decreased as the aspect ratio of the gain strip $L_p/W_p$ increased. Similar results were obtained when we varied $W_p$ and kept $L_p$ constant. These results indicated that the output from a two-particle cavity laser cannot be simply regarded as scattering of a plane wave by a single particle even if $L_c \gg \lambda$. It relied on both the geometry of the scatterers and the shape of the gain region. The directionality  of the lasing output is a consequence of gain guiding.  

Although in the dilute suspension of particles there was probably only one particle at the tip of the excited cone, there were more particles at the cone base whose dimension exceeded the average distance between the particles. For example, when the particle density $\rho = 5 \times 10^8$ cm$^{-3}$ and pump spot radius $\sim$ 20 $\mu$m, there were typically ten particles near the base of the excited cone. To simulate this situation, we placed ten scatterers randomly near one end of the gain strip and only one scatterer at the other end. The lasing peaks were almost equally spaced in frequency, with the spacing close to that with only two scatterers in the strip, one on either end. It suggested that the feedback from the ten scatterers near one end of the gain strip was equivalent to that from one located somewhere close to this end, as far as the lasing frequencies were concerned.  

One question we raised at the end of the previous section is why the feedback from the particles near the two ends of the excited cone dominated over that from the particles inside the cone. One possible explanation would be that light backscattered by the particles near one end of the cone experienced the most amplification as it traveled the longest path within the gain volume to the other end of the cone. If this were the reason, it implied the lasing modes would differ from the quasimodes of the passive system (without gain or loss). To check this conjecture, we randomly placed four cylinders inside the gain strip (8 $\mu$m $\times$ 4 $\mu$m) in addition to the two at the ends, and compared the lasing modes to the quasimodes  of the passive system.  The quasimodes were calculated  with the  multipole  method \cite{ara,botten}. The field around each cylinder was expanded in a Fourier-Bessel series of regular and outgoing cylindrical harmonic functions.   The Rayleigh identity related the regular part of the field at a particular cylinder to the waves sourced at all other scatterers. We found the quasimodes by searching in the complex wavelength plane for the poles of the scattering operator. 

Our calculations revealed that despite the presence of additional scatterers, lasing still occurred in the direction parallel to the strip. Moreover, the lasing modes corresponded to the quasimodes with quality factor relatively high among all the quasimodes within the gain spectrum. Table 1 listed the frequencies $\nu$ and quality factors $Q$ of several lasing modes and the corresponding quasimodes. The gain spectrum was centered at 750 THz with a width of about 281 THz.  The slight frequency shift of the lasing modes with respect to the quasimodes was due to the gain pulling effect. Figure 10 shows the intensity distributions of the lasing mode with $\nu =764.9$ THz and the quasimode with $\nu =764.5$ THz. It is evident that the lasing mode profile within the gain strip is nearly identical to that of the quasimode.  We checked several lasing modes and obtained the same result. Therefore, the lasing modes were almost the same as the quasimodes in the presence of uniform gain. 

In Table 1, we also listed the frequencies and quality factors of the quasimodes in the two-particle case (without the four particles in the middle). Majority of the quasimodes in the system of six particles have frequencies similar to those of two particles. This comparison suggests in the weakly scattering system most quasimodes with relatively high $Q$ are formed mainly by the feedback from the particles near the system boundary. The feedback from the particles in the interior of the system may slightly increase the quality factor or shift the mode frequency (see, e.g., the mode at $\nu$ = 742.2 THz). However, the feedback from these particles may also be destructive and reduce the quality factor. As a result, the mode at $\nu =  736.6$ THz would not lase.  Because most quasimodes with relatively high $Q$ have frequencies similar to those with only the two particles furtherest apart, they, as well as the lasing modes, tend to be equally spaced in frequency. 

In the numerical simulation we did not place the particles outside the gain strip, because experimentally the feedback from those paticles is suppressed by reabsorption. Thus the effective system size is reduced, as shown in our previous calculation \cite{yamilov}.

\section{Discussion and Conclusion}

Our experimental studies demonstrated lasing with field feedback in weakly scattering samples. The focused pump beam created a cone-shaped gain volume in the dye solution containing a small amount of nanoparticles. The cone length was determined by the absorption of dye molecules since optical scattering was much weaker than absorption. When the scattering mean free path exceeded the size of gain volume, lasing oscillation built up in the direction of strongest amplification, i.e., the direction in which the gain volume was most extended. This behavior was similar to that of amplified spontaneous emission in the weak scattering regime \cite{wiersma1}. The fundamental difference from ASE was, however, the existence of feedback that originated from the backscattered light. As pointed out by Kumar and coworkers \cite{kumar}, the statistically rare sub-mean-free-path scattering could be made effective by strong amplification. In our experiment, the extreme weakness of feedback was compensated by high optical gain due to intense pumping. The interference of the backscattered light was greatly enhanced, leading to coherent and resonant feedback for lasing. 

It is important to note that the discrete lasing peaks were distinct from the stochastic ASE spikes. The latter originated from random spontaneous emission, which was significantly amplified in the presence of large gain. Such spectral fluctuations also existed in ASE from the homogeneous media (without scattering). Hence, scattering was not indispensable to the existence of ASE spikes, although the fluctuations could be enhanced by scattering which stretched the path length of photons inside the gain volume. In contrast, the presence of scatterers in the gain media was essential to the emergence of lasing peaks, indicating the lasing process relied on the feedback supplied by scattering.   

There have been many theoretical investigations of lasing modes in the weak scattering regime. They can be either extended modes \cite{pinheiro} or anomalously localized modes \cite{apalkov}. If only a small part of the random medium is pumped, the anomalously localized modes that locate inside the pumped region experience more gain than the extended modes that spread over the entire system. Therefore, the anomalously localized modes may have lower lasing threshold. However, the anomalously localized states are extremely rare in the weakly scattering samples. An alternative mechanism for spatial localization of lasing modes is absorption of emitted light outside the pumped region \cite{yamilov}. The reabsorption of emission suppresses the feedback from the unpumped part of the random system and effectively reduces the system size. The lasing modes are therefore drastically different from the quasimode of the passive system (without gain or absorption). Even if all the quasimodes of the passive system are extended across the entire system, the lasing modes are still confined in the vicinity of the gain volume.  

The numerical simulations in section III also illustrate that the lasing modes are nearly identical to the quasimodes of the reduced system. The quasimodes are formed by distributed feedback from all the particles inside the reduced system. The conventional distributed feedback (DFB) lasers, made of periodic structures, operate either in the over-coupling regime or the under-coupling regime \cite{kogelnik}. The random lasers, which can  be regarded as randomly distributed feedback lasers, also have these two regimes of operation. In the under-coupling regime the quasimodes are formed {\it mainly} by the feedback from the scatterers near the system boundary, while in the over-coupling regime the feedback from the scatterers inside the system becomes important. Thus, the quasimodes of a under-coupled system, especially the ones with relatively high quality, have almost regular frequency spacing. Note that the feedback from the scatterers inside the system is weak but not negligible, e.g., it may cause a slight shift of mode frequency or modification of the quality factor. In our experiment with the dilute suspensions of particles, the random lasing was in the under-coupling regime as a result of weak scattering and small size of the reduced system. Therefore, the dominant feedback from the particles near the cone ends resulted in nearly constant frequency spacing of the lasing peaks, which scaled inversely with the cone length. Due to weak feedback from the particles inside the cone, the lasing modes were not exactly equally spaced in frequency and some modes failed to lase as their quality factors were reduced. In the previous studies, e.g. Refs.\cite{frolov,frolov1,frolov2,caoPRL99}, the strong scattering or large system size make the random laser operate in  the over-coupling regime. Thus the lasing peaks are randomly spaced.      
 
The results of our studies not only illustrated the  physical mechanism of random lasing in the weak scattering regime, but  also  demonstrated the possibility of  controlling the frequencies and output directionality of random lasers by varying the pumping geometry, the scattering mean free path, and the absorption length at both excitation and emission wavelengths. Such control is important to the application of random lasers. 

The authors acknowledge Dr. Christian Vanneste for numerous stimulating discussions and critical reading of this manuscript. This work was supported by the National Science Foundation under the grant no. ECS-0244457 and ECS-0601249, and by the Australian Research Council under its Discovery Grants and Centres of Excellence Programs.

\newpage

\noindent Table 1. Frequencies $\nu$ (THz) and quality factors $Q$ of lasing modes and quasimodes.
\newline

\begin{tabular}{|c|c|c|c|c|} \hline
6-particle  & 6-particle & 6-particle & 2-particle &  2-particle  \\ \hline
lasing $\nu$ & quasimode $\nu$ & quasimode $Q$ & quasimode $\nu$ & quasimode $Q$ \\ \hline
 715.6 & 712.1 & 28.6 & 712.4 & 27.2 \\ \hline
	     & 736.6 & 18.7 & 730.5 & 28.2 \\ \hline
 744.0 & 742.2 & 31.0 & 748.5 & 29.3 \\ \hline
 764.9 & 764.5 & 29.4 & 766.5 & 30.5 \\ \hline
 782.9 & 785.1 & 37.7 & 784.1 & 31.8 \\ \hline
 806.9 & 811.7 & 38.5 & 801.7 & 32.8 \\ \hline
\end{tabular} 

\newpage

\begin{figure}[h]\centerline{\scalebox{1.0}{\includegraphics{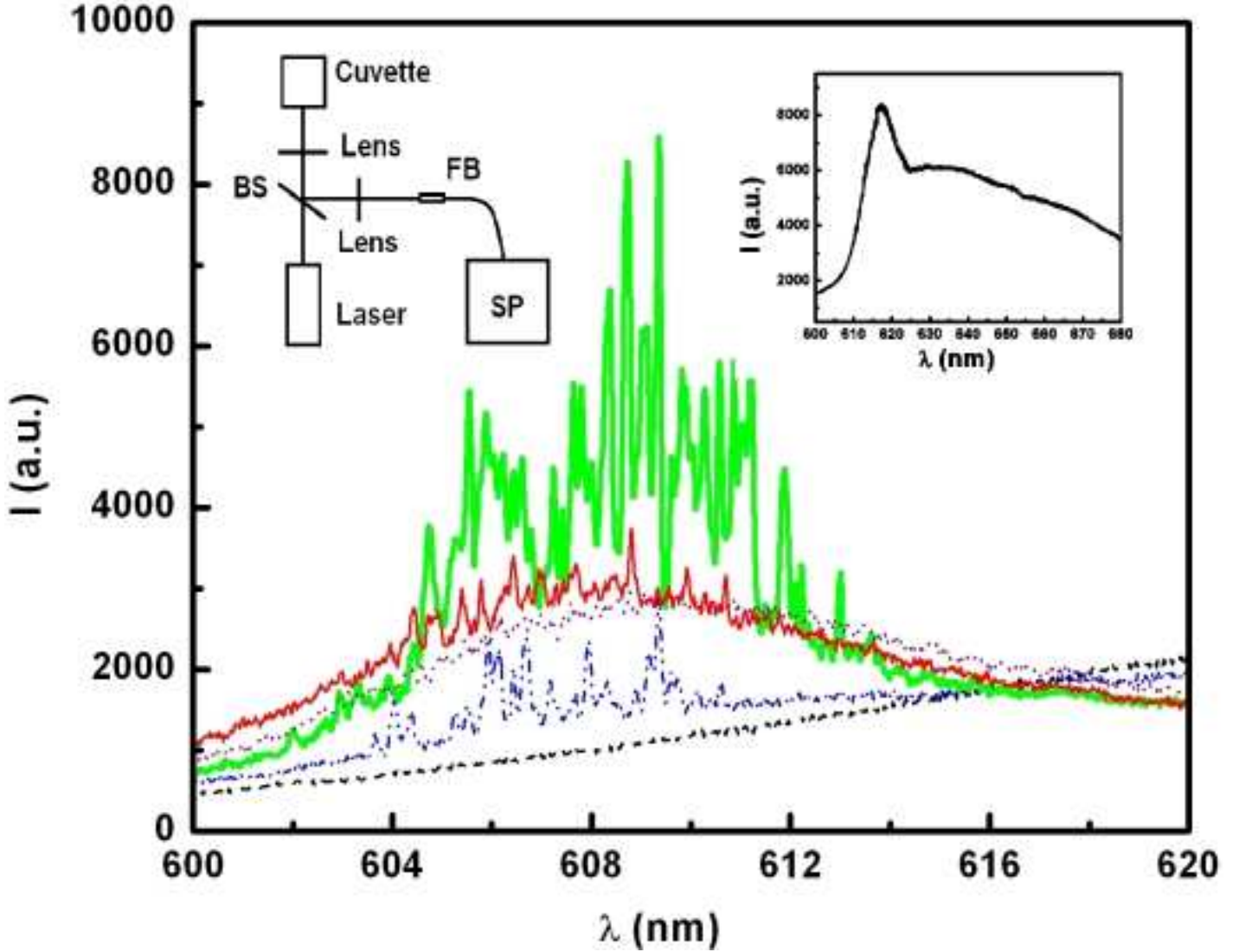}}}
\caption{Spectra of emission from DEG solutions of Rhodamine 640 (5 mM) and TiO$_2$ particles. The particle densities are (1)$\rho$ = 0 (dotted curve), (2) $1.87\times10^{8}$ cm$^{-3}$ (thin solid curve), (3) $1.87\times10^{9}$ cm$^{-3}$(thick solid curve), (4) $1.3\times10^{10}$ cm$^{-3}$(dot-dot-dashed curve), (5) $5.0\times10^{10}$ cm$^{-3}$ (dashed curve). All spectra were taken at the same pump pulse energy 0.4 $\mu$J. Each spectrum was integrated over 25 shots. Left inset is our experimental setup, BS: beam splitter, SP: spectrometer, FB: fiber bundle. Right inset is the emission spectrum shows the ASE peak. $\rho = 5.0\times10^{10}$ cm$^{-3}$. The pump pulse energy is 1.2 $\mu$J.}
\end{figure}

\newpage

\begin{figure}[h]\centerline{\scalebox{1.0}{\includegraphics{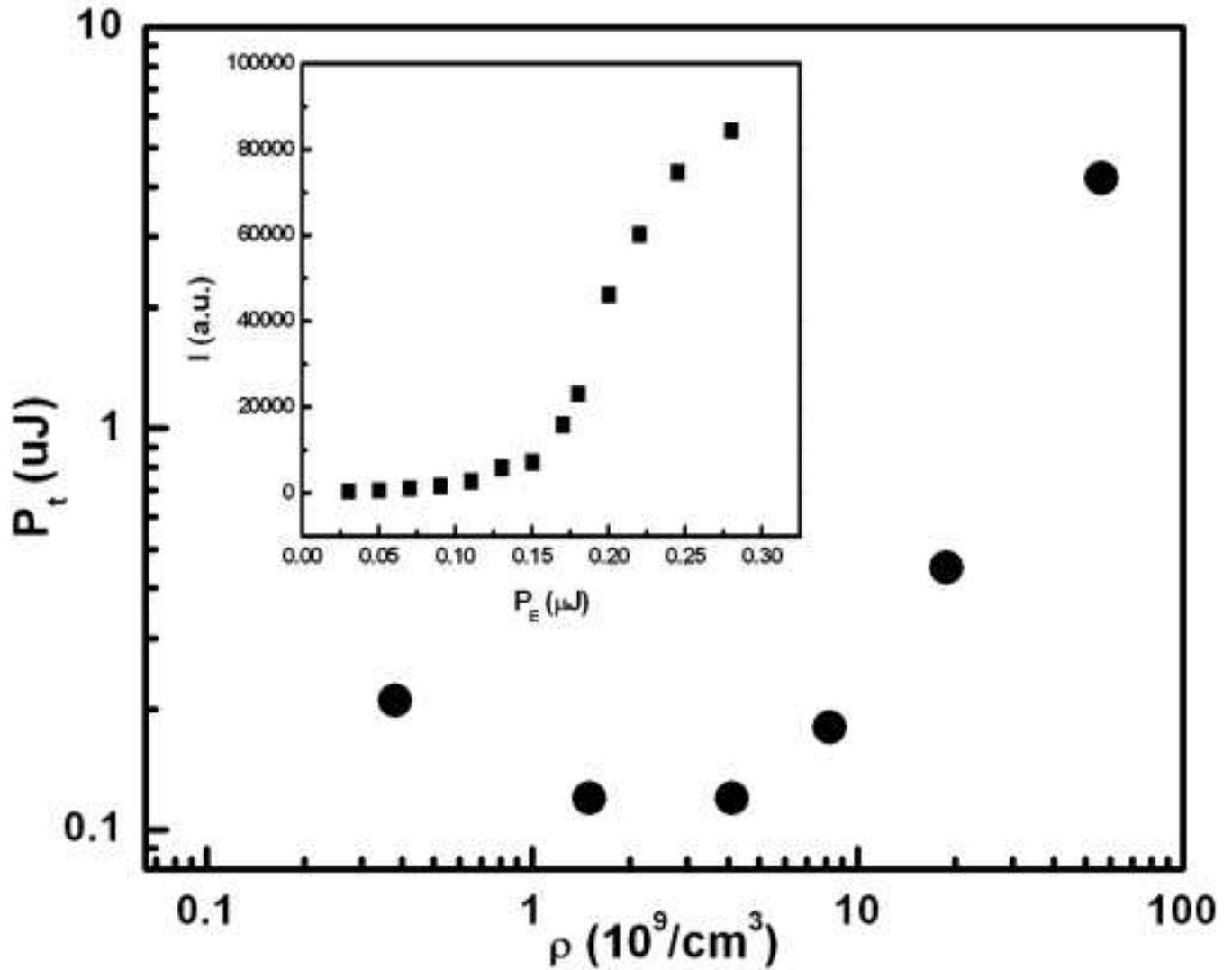}}}
\caption{The threshold pump pulse energy $P_{t}$ as a function of the TiO$_2$ particle density$\rho$. The concentration of Rhodamine 640 in DEG is 5 mM. The inset is a plot of the emission intensity $I$ versus the pump pulse energy $P_E$ for the suspension with $\rho = 3.0\times10^{9}$ cm$^{-3}$.}
\end{figure}
\newpage

\begin{figure}[h]\centerline{\scalebox{1.0}{\includegraphics{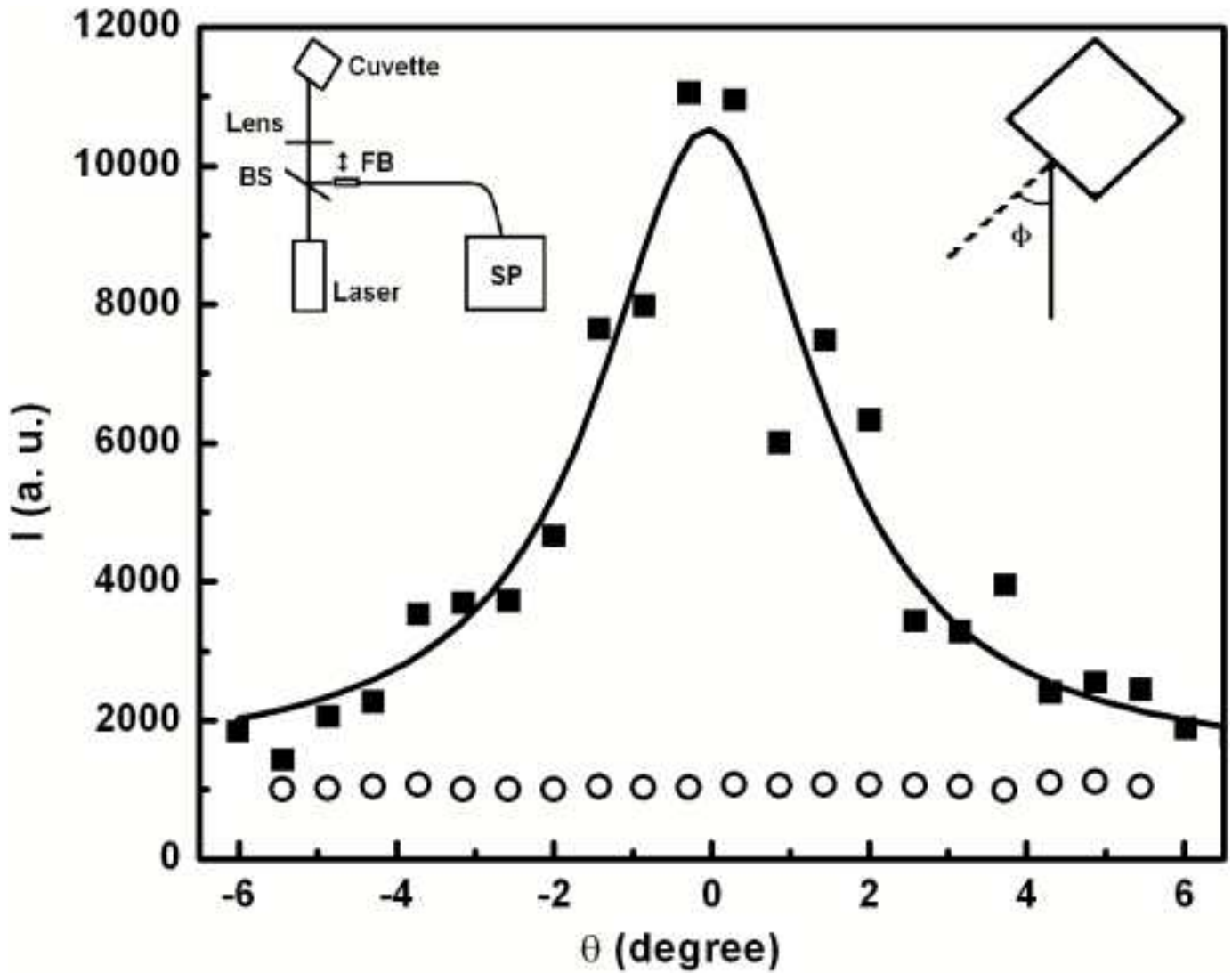}}}
\caption{The angular distribution of output emission intensity from DEG solutions of 5 mM Rhodamine 640 and $3.0\times10^{9}$ cm$^{-3}$ (solid square) or $5.0\times10^{10}$ cm$^{-3}$ (open circle) TiO$_{2}$ particles. The solid line is a Gaussian fit of output laser beam. $\theta = 0$ corresponds to the backward direction of pump beam. Left inset is a sketch of experimental setup, BS: beam splitter, SP: spectrometer, FB: fiber bundle. Right inset shows the angle $\phi$ between the pump beam and the normal to the front window of the cuvette.}
\end{figure}
\newpage

\begin{figure}[h]\centerline{\scalebox{1.0}{\includegraphics{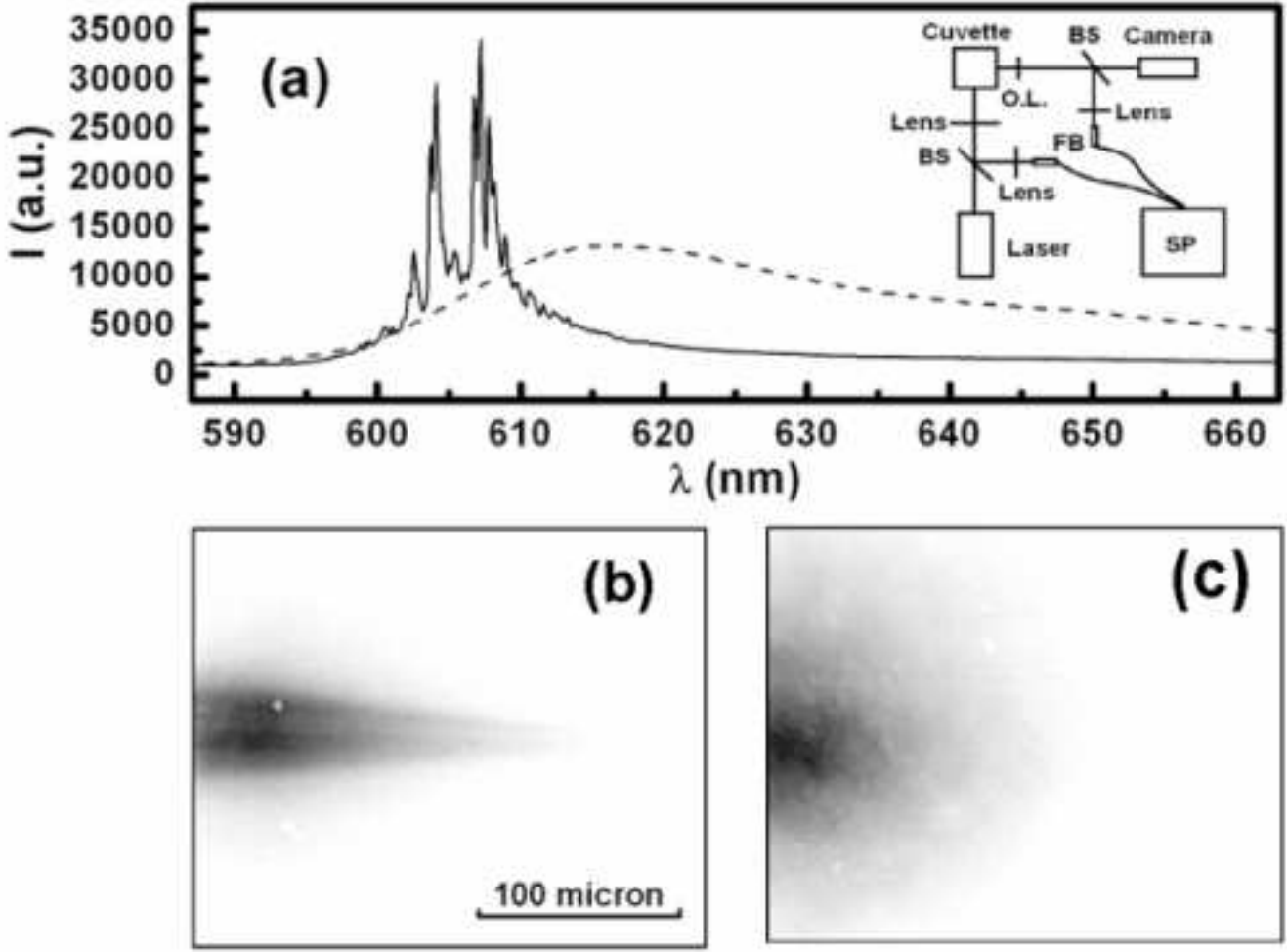}}}
\caption{(a): Spectra of emission through the side window (dashed curve) and front window (solid curve) of the cuvette. The inset is a sketch of the experimental setup, BS: beam splitter, SP: spectrometer, FB: fiber bundle. (b): Side image of excited region in DEG solution of 5 mM Rhodamine 640 and $3.0\times10^{9}$ cm$^{-3}$ TiO$_2$ particles. The pump pulse energy is 0.2 $\mu$J. (c): Side image of excited region in DEG solution of 5 mM Rhodamine 640 and $5.0\times10^{10}$ cm$^{-3}$ TiO$_2$ particles. The pump pulse energy is 1.2 $\mu$J.}
\end{figure}
\newpage

\begin{figure}[h]\centerline{\scalebox{1.0}{\includegraphics{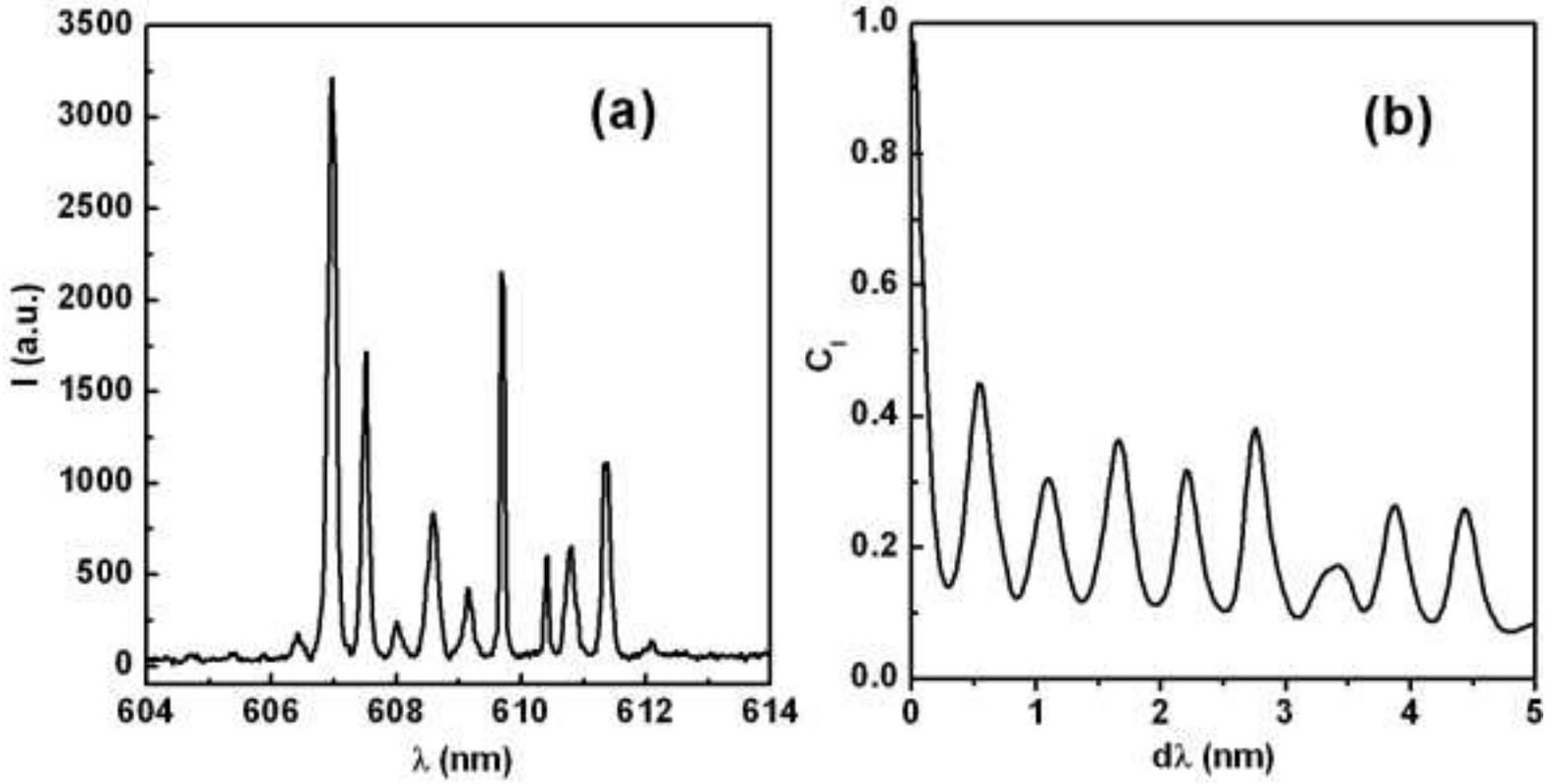}}}
\caption{(a): A single shot emission spectrum from DEG solution of 5 mM Rhodamine 640 and $1.87\times10^{9}$ cm$^{-3}$ TiO$_{2}$ particles. The pump pulse energy is 0.2 $\mu$J pumping. (b): Spectral correlation function computed for the spectrum in (a).}
\end{figure}
\newpage

\begin{figure}[h]\centerline{\scalebox{1.0}{\includegraphics{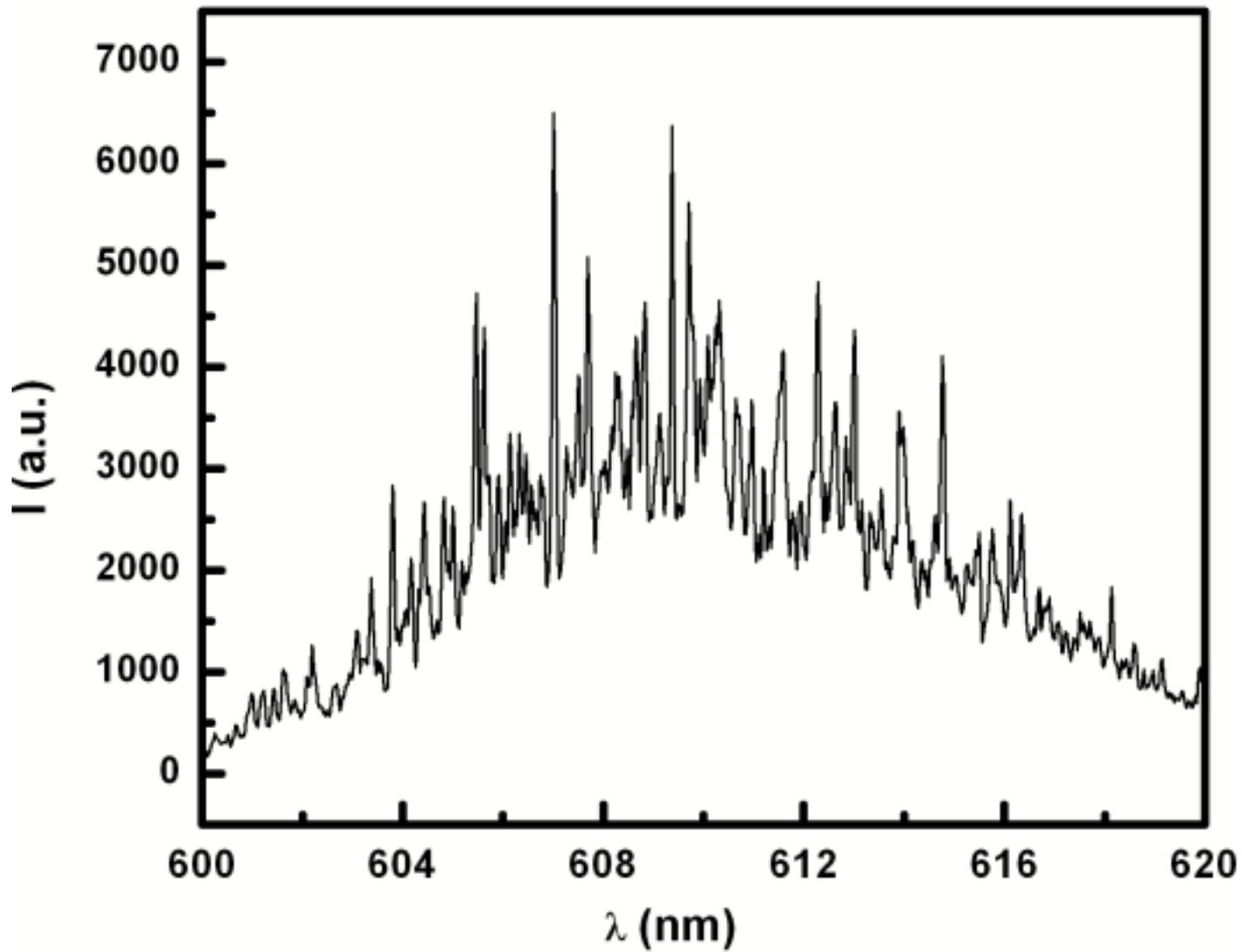}}}
\caption{A single shot emission spectrum from the DEG solution of 5 mM Rhodamine 640 without any particles. The pump pulse energy is 0.3 $\mu$J pumping.}
\end{figure}
\newpage

\begin{figure}[h]\centerline{\scalebox{1.0}{\includegraphics{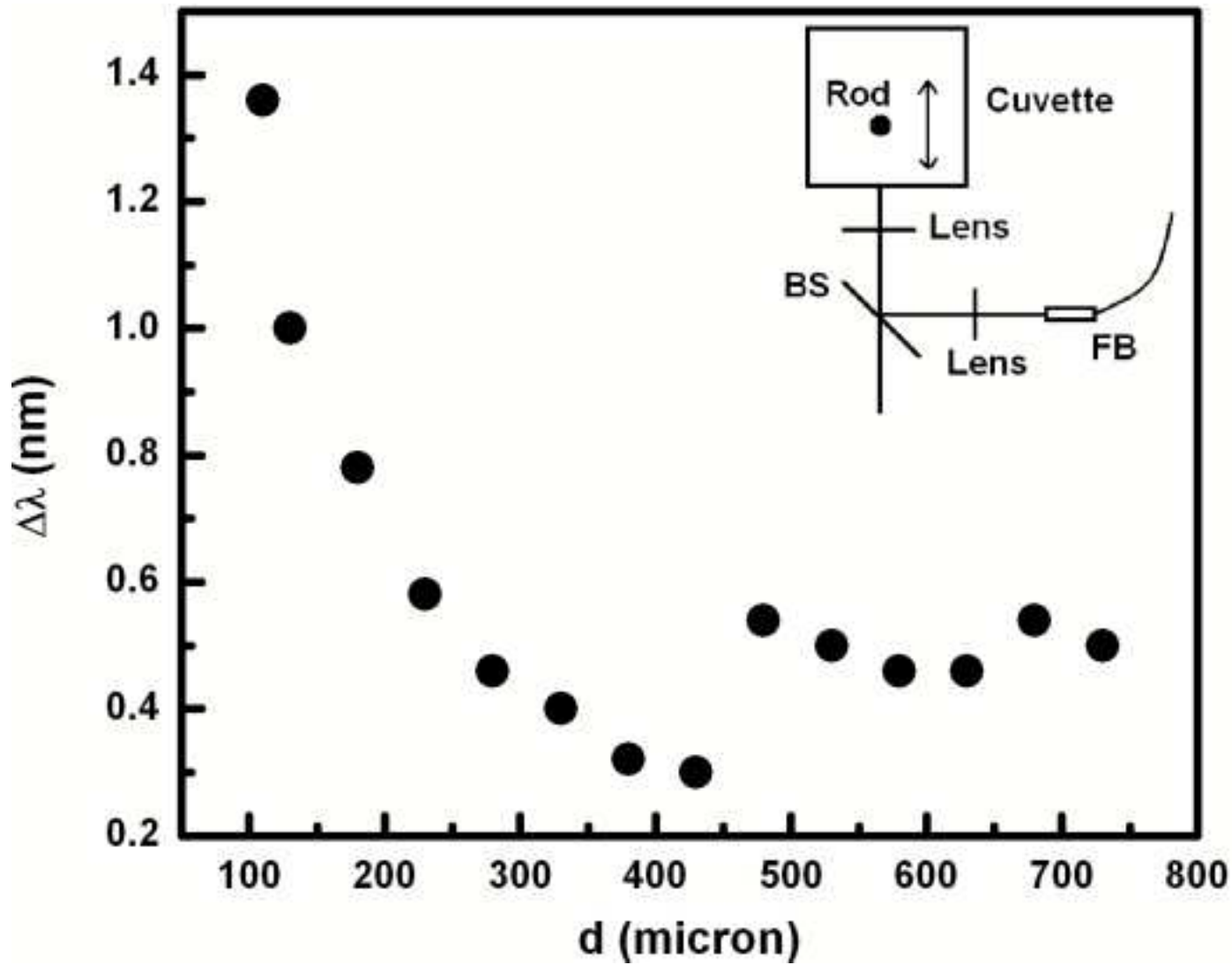}}}
\caption{The wavelength spacing $\Delta \lambda$ of lasing modes as a function of the distance $d$ between the metallic rod and the cuvette front window. The concentration of Rhodamine 640 in DEG is 5 mM, and the TiO$_2$ particle density is $3.0\times10^{9}$ cm$^{-3}$. The pump pulse energy is  0.2 $\mu$J. The inset is a sketch of experimental setup. BS: beam splitter, FB: fiber bundle.}
\end{figure}
\newpage

\begin{figure}[h]\centerline{\scalebox{1.0}{\includegraphics{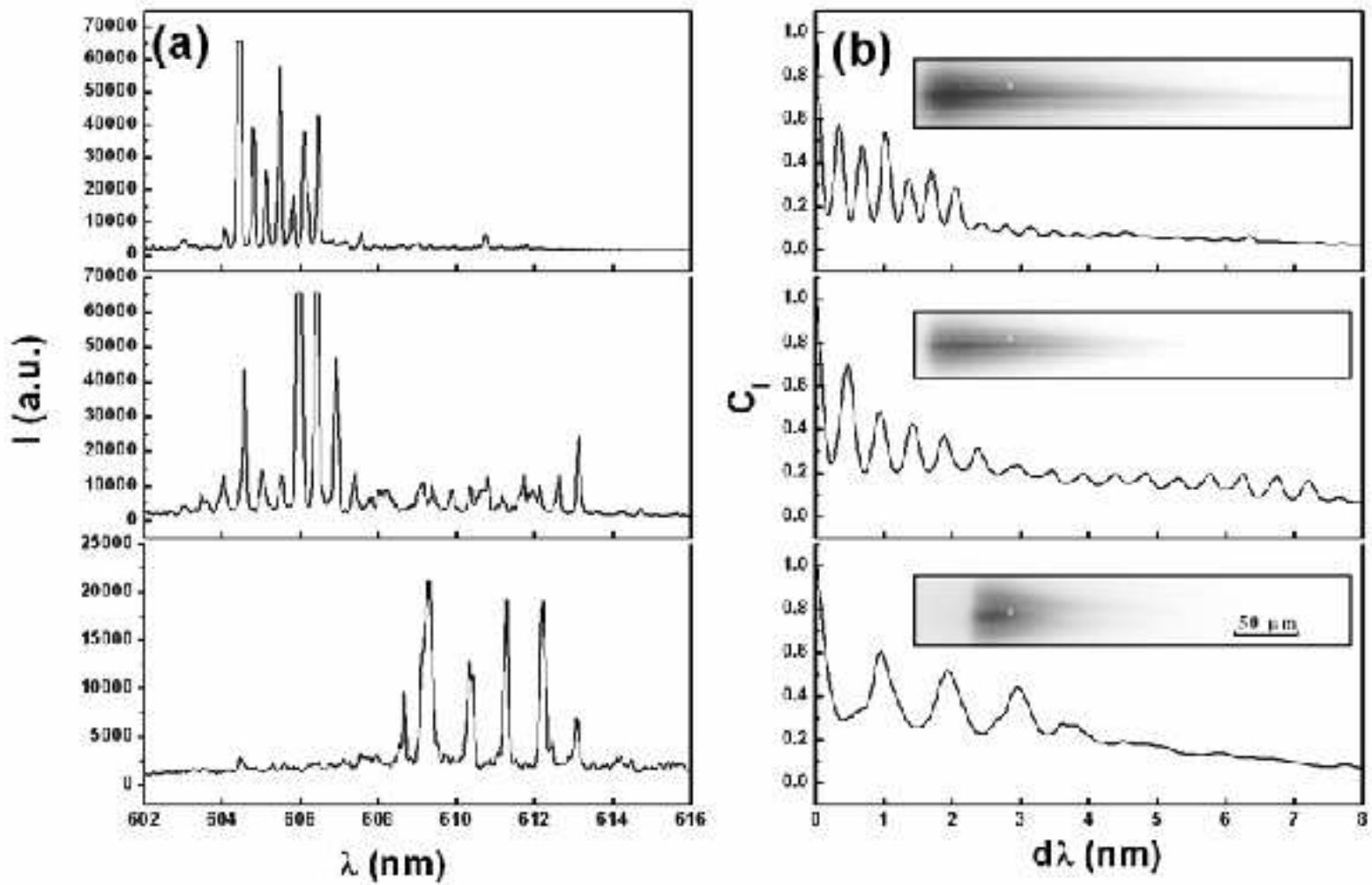}}}
\caption{(a): Single shot emission spectra from colloidal solution of $3.0\times10^{9}$ cm$^{-3}$ TiO$_2$ particles. The molarity of Rhodamine 640 in DEG is (from top to bottom) 3 mM, 5 mM, and 10 mM.  (b) Spectral correlation functions of the single shot emission spectra in (a). Insets are side images of pumped region for in each solution.}
\end{figure}
\newpage

\begin{figure}[h]\centerline{\scalebox{1.0}{\includegraphics{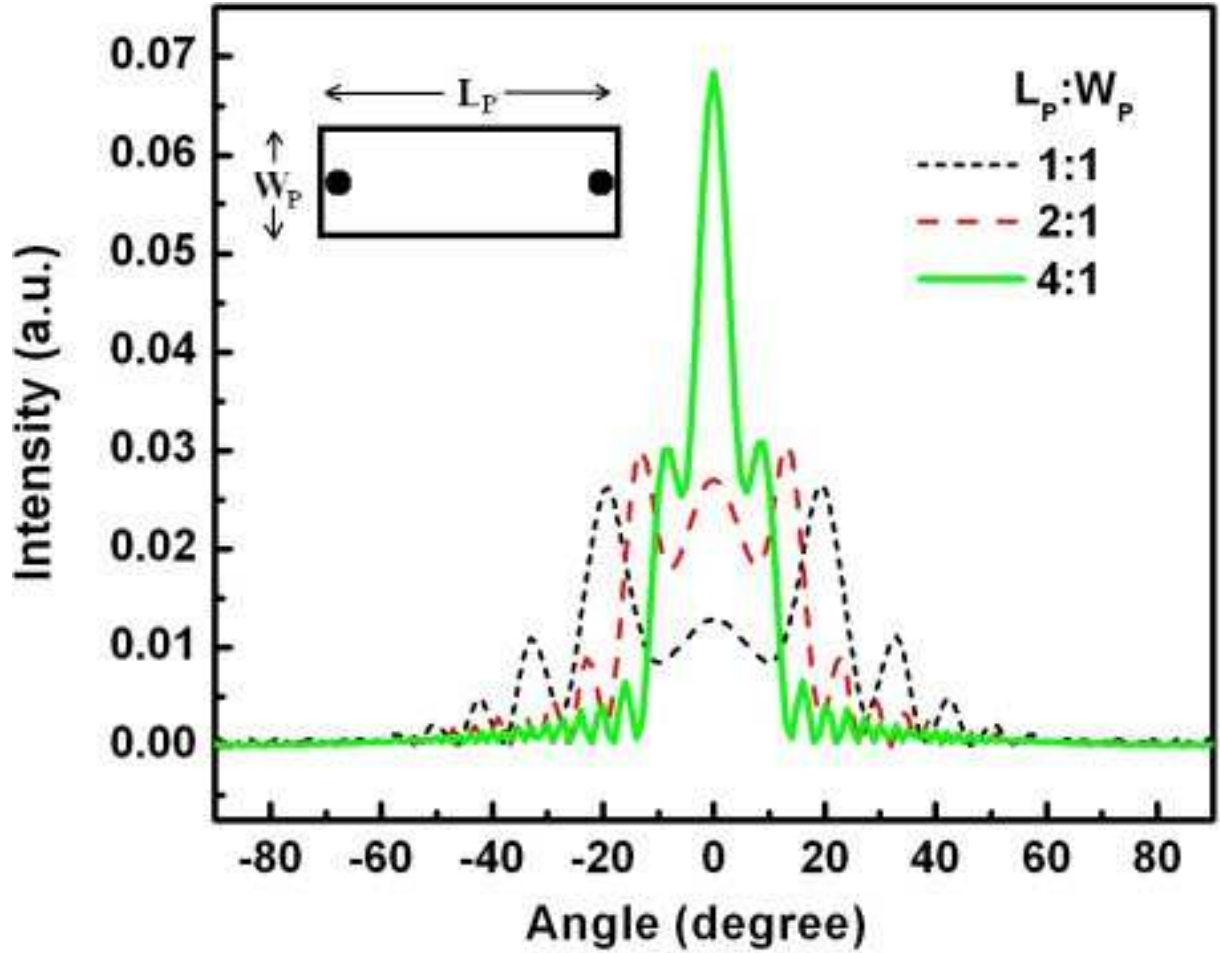}}}
\caption{Far-field intensity of laser emission as a function of polar angle. The zero degree corresponds to the direction parallel to the gain strip. The width of gain strip $W_p$ was fixed  at  4  $\mu$m. The strip length $L_p$ =  4 $\mu$m (dotted curve), 8  $\mu$m (dashed curve), and 16 $\mu$m (solid curve). The inset is a sketch of the geometry of the 2D system in our numerical simulation. Two cylinders of radius 100 nm are located on both ends of the gain strip. }
\end{figure}
\newpage

\begin{figure}[h]\centerline{\scalebox{1.0}{\includegraphics{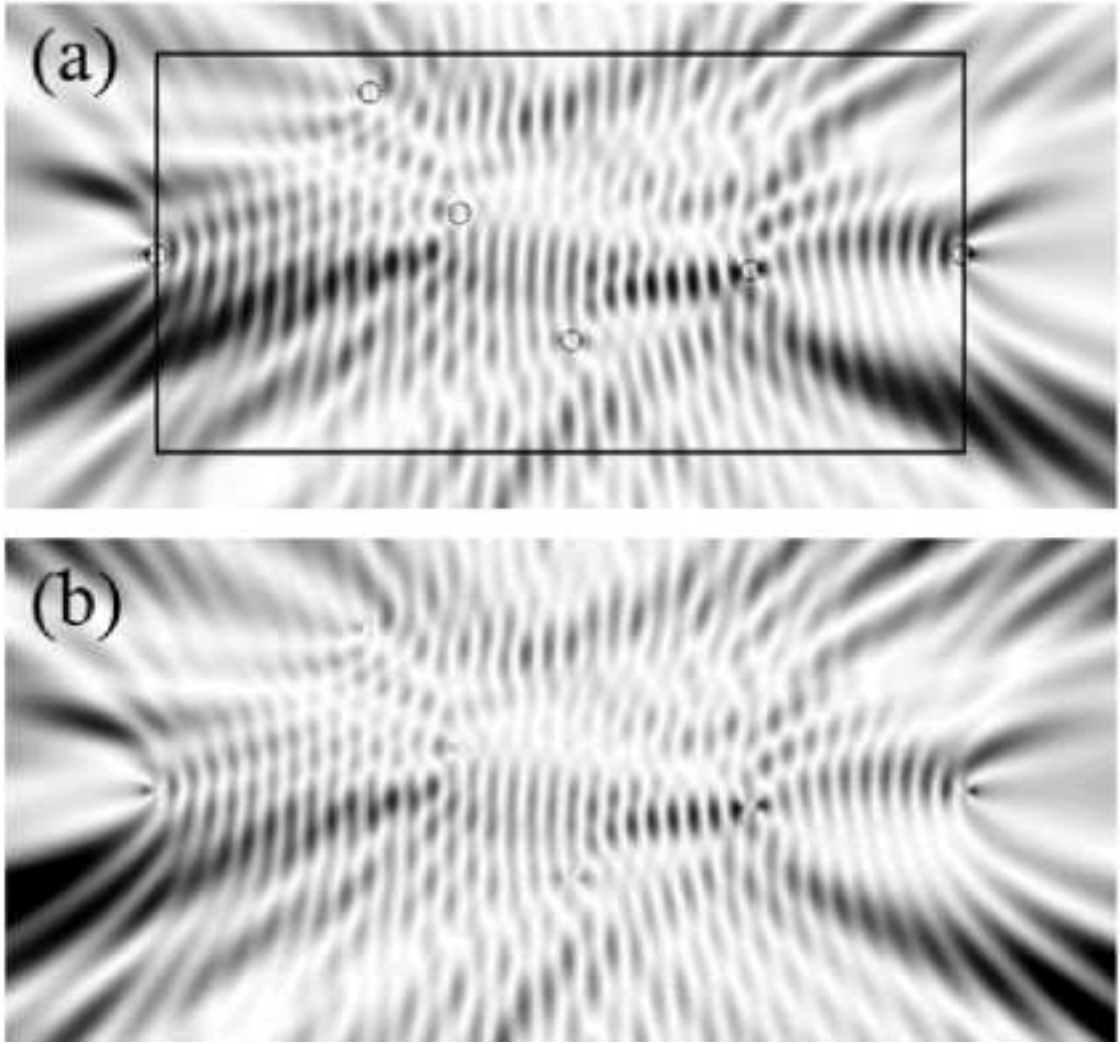}}}
\caption{(a) Spatial intensity distribution of a lasing mode at $\nu$ = 764.9 THz in the 2D system of six dielectric cylinders inside a gain strip 8 $\mu$m  $\times$ 4 $\mu$m. The circles represent the cylinders. The rectangle marks the boundary of the gain strip. (b) Spatial intensity distribution of quasimode at $\nu$ = 764.5 THz in the same system as (a) but without gain.}
\end{figure}

\end{document}